\documentclass[aps,pra,twocolumn,superscriptaddress,longbibliography]{revtex4-1}
\usepackage{graphicx}

\usepackage{amssymb,amsfonts,amsmath}
\begin{document}

%%%%%%%%%%%%%%%%%%%%%%%%%%%%%%

%% For titles, only capitalize the first letter
%% \title{Almost sharp fronts for the surface quasi-geostrophic equation}

\title{Morphodynamics Facilitate Cancer Cells to Navigate 3D Extracellular Matrix}

\author{Christopher Z. Eddy}
\affiliation{Department of Physics, Oregon State University, Corvallis OR, 97331}
\author{Helena Raposo}
\affiliation{Department of Chemical, Biological, and Environmental Engineering, Oregon State University, Corvallis OR, 97331}
\author{Aayushi Manchanda}
\affiliation{Molecular and Cellular Biology Program, Oregon State University, Corvallis OR, 97331}
\author{Ryan Wong}
\affiliation{Department of Physics, Oregon State University, Corvallis OR, 97331}
\author{Fuxin Li}
\affiliation{School of Electrical Engineering and Computer Science, Oregon State University, Corvallis OR, 97331}
\author{Bo Sun}
\thanks{Correspondence to sunb@onid.orst.edu}
\affiliation{Department of Physics, Oregon State University, Corvallis OR, 97331}

%%%%%%%%%%%%%%%%%%%%%%%%%%%%%%%%%%%%%%%%%%%%%%%%%%%%%%%%%%%%%%%%

\begin{abstract}
Cell shape is linked to cell function. The significance of cell morphodynamics, namely the temporal fluctuation of cell shape, is much less understood. Here we study the morphodynamics of MDA-MB-231 cells in type I collagen extracellular matrix (ECM). We systematically vary ECM physical properties by tuning collagen concentrations, alignment, and gelation temperatures. We find that morphodynamics of 3D migrating cells are externally controlled by ECM mechanics and internally modulated by Rho/ROCK-signaling. We employ machine learning to classify cell shape into four different morphological phenotypes, each corresponding to a distinct migration mode. As a result, we map cell morphodynamics at mesoscale into the temporal evolution of morphological phenotypes. We characterize the mesoscale dynamics including occurrence probability, dwell time and transition matrix at varying ECM conditions, which demonstrate the complex phenotype landscape and optimal pathways for phenotype transitions. In light of the mesoscale dynamics, we show that 3D cancer cell motility is a hidden Markov process whereby the step size distributions of cell migration are coupled with simultaneous cell morphodynamics. Morphological phenotype transitions also facilitate cancer cells to navigate non-uniform ECM such as traversing the interface between matrices of two distinct microstructures. In conclusion, we demonstrate that 3D migrating cancer cells exhibit rich morphodynamics that is controlled by ECM mechanics, Rho/ROCK-signaling, and regulate cell motility. Our results pave the way to the functional understanding and mechanical programming of cell morphodynamics as a route to predict and control 3D cell motility.

\end{abstract}

\maketitle

% \section{Significance Statement}
% Cell shape has been an important biomarker since the beginning of cell
% biology. While most studies focus on the static cell morphology, we
% examine the spontaneous shape fluctuations, namely the morphodynamics,
% of 3D migrating cancer cells. By employing machine learning techniques we map cell morphodynamics into transitions
% between different morphological phenotypes which are associated with
% distinct migration modes. We find cells rapidly switch morphological
% phenotypes and the transitions are governed by the ECM
% mechanics and cell mechanotransduction. Phenotype transition also facilitates the invasion of
% cancer cells. Taken together, our study reveal the complex landscape of morphological phenotypes. The results pave the way to establish
% morphodynamics as single cell trait to facilitate the prediction and programming of cell dynamics.
\section{Introduction}
\label{sec:Intro}
Shape defines the cell. In the 1677 book \textit{Micrographia}, Robert Hooke showed sections within a herbaceous plant under a microscope. The shape of those sections resembles cells in a monastery, so he named the structures cells \cite{Hooke1665}. Many breakthroughs followed Hooke's discovery, from the cell theory of Schwann and Schleiden, to the theory of tissue formation by Remak, Virchow and Kolliker, and the theory of cellularpathologie by Virchow, all of which are inspired by observations of cell shapes, or morphology in general \cite{Mazzarello1999,Mayr1982}.

In our modern view cell shape is determined by cell function \cite{Walter2014,Ingber1994}. A nerve cell has long branched protrusions for communication with other neurons; while the cuboidal shape of epithelial cells allow them to tile the surface of organs. Loss of characteristic shape, on the other hand, is associated with functional abnormality. Thus morphological characterization has been an important tool for diagnosis such as in red blood cell disease \cite{Diez2010}, neurological disease \cite{Serrano2011}, and cancer \cite{Bakal2013,Wu2015}. More recently, cell shape analysis is boosted by techniques from computer vision. As a result, it becomes possible to obtain high content information of cellular states from morphological data alone \cite{Perrimon2007,Gerlich2010,Wu2015,Lam2017,Carpenter2017}.

While most research focuses on the static cell morphology, the dynamic fluctuation of cell shape is much less understood \cite{Huber2013,Bathe2016}. However, shape fluctuation -- namely morphodynamics, is of central importance for dynamic cellular functions. The abnormal diffusion of small protrusions - microvilli - on the surface of a T cell allows the T cell to efficiently scan antigen-presenting surfaces \cite{Caieaal3118}. For a migrating cancer cell, morphodynamics drives the motility of the cell in many ways similar to our body frame movements that enable swimming. In fact, just as there are different swimming styles, cancer cells have been observed to execute multiple programs -- migration modes -- during invasion in 3D tissue space \cite{Konstantopoulos2017}. Each mode has distinct signatures of morphology and morphodynamics, and are usually classified based on cell morphology as filopodial, lamellipodial, lobopodial, blebbing, and actin-enriched leading edge \cite{Yamada2012commentary}. Cancer cell migration modes is controlled by intracellular signaling such as the Rho-ROCK-myosin pathways \cite{Marshall2008rac,Marshall2005rho}, and extracellular factors such as the elasticity, and degradability of the extracellular matrix (ECM) \cite{Marshall2010plasticity,Yamada2012commentary}. The ability of a cancer cell to switch between migration modes is important for tumor prognosis. Many therapies, such as MMP inhibitors that target a particular mode of cell motility, fail to stop tumor metastasis largely because cells take other available migration programs \cite{Zucker2003,Friedl2003MMP_inhibition_transition}.

In this paper, we study the morphodynamics of MDA-MB-231 cells, a highly invasive human breast cancer cell line, in 3D collagen matrices. We devise machine learning techniques to classify cell shapes into morphological phenotypes that correspond to known migration modes. This approach provides a mesoscale mapping of cell morphodynamics into transitions among morphological phenotypes. We find individual cells are capable of rapidly sampling multiple morphological phenotypes, implying spontaneous migration mode transitions. We find ECM mechanics coupled with cell mechanosensing pathways regulate the stability and transition rates between morphological phenotypes. We also find that such transitions facilitate cancer cells to navigate ECM with inherent structural and mechanical heterogeneity. Our results reveal 3D cancer cell migration as a hidden Markov process and morphodynamics contribute to the changes of motility by ECM physical cues.

\begin{figure}[h]
 \centering \includegraphics[width=0.99\columnwidth]{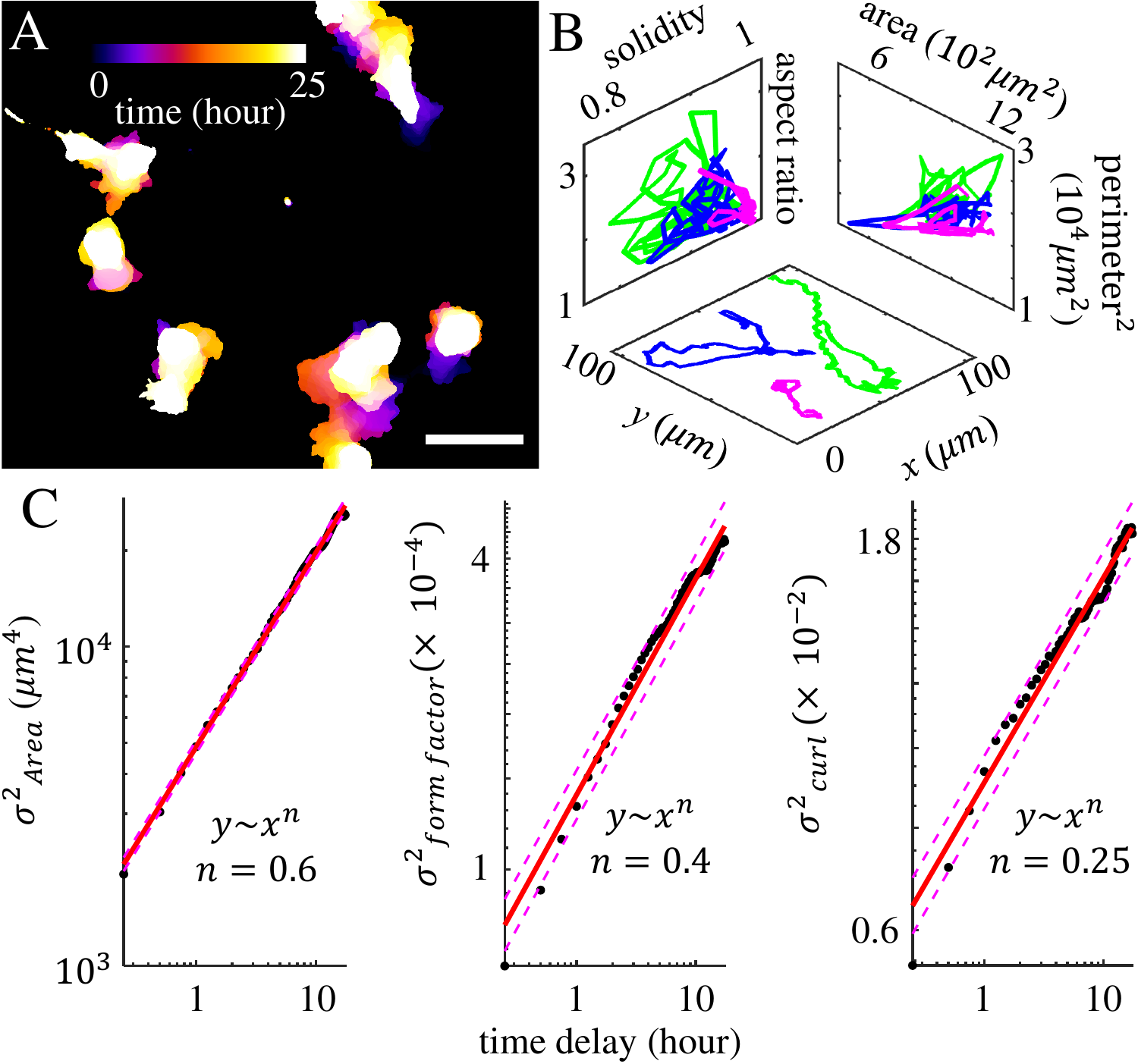}
 \caption{Three-dimensional migration of MDA-MB-231 cells show
   significant cell shape fluctuation. (A) A typical
   time lapse recording of 25 hours is projected onto a single image with colors
   representing time. (B) The real space (x-y plane), and shape fluctuations of 3 cells shown in (A). (C) The mean square displacement ($\sigma^2$) of selected cell geometric measures. Dots: experimental measurements. Solid lines: linear fit. Dashed lines: 95\% prediction interval. Here the form factor is defined as $\text{perimeter}^2/\text{area}$. Curl is defined as the ratio between the major axis length and skeletonized contour length. This figure is prepared with Mathlab R2020a (www.mathworks.com) and ImageJ (https://imagej.net).}
\label{fig1}
\end{figure}
\section{Results}
\label{sec:results}

We find 3D migrating cancer cells demonstrate rapid shape fluctuations (Fig. \ref{fig1}(A-B)). In order to quantify the cell morphodynamics, we take time-lapse fluorescent images of MDA-MB-231 cells migrating in collagen matrices. The GFP-labeled cells typically stay within the focal depth of the objective lens (20X, NA 0.7) for 10-20 hours, while we obtain 2D cell images at a rate of 4 frames per hour (see \textit{SI Appendix} section S1). After binarization and segmentation, we compute a total of twenty-one geometric measures which collectively quantify the shape of a cell (see \textit{SI Appendix} section S2). These geometric measures characterize cell size (such as area and perimeter), deviation from circle (such as aspect ratio and form factor), surface topography (such as solidity), and backbone curvature (such as curl -- the ratio between the major axis length and skeletonized contour length).

The morphodynamics of a cell manifests itself as a random walk in the geometric shape space concurrent with its motility in the 3D matrix (Fig. \ref{fig1}). However, unlike the real space motility that is slightly superdiffusive \cite{Sun20133d}, we find cell morphodynamics is subdiffusive in the geometric shape space (Fig. \ref{fig1}C and \textit{SI Appendix} section S3). The subdiffusivity suggests physical barriers that are present both intrinsic to the cells and from the 3D ECM. Indeed, we find cells moving on 2D surface exhibit faster shape fluctuations than cells embedded in 3D ECM. Still, on flat surfaces cells show subdiffusive random walks in the geometric shape space and superdiffusive walks in real space (SI Appendix section S3).

\begin{figure}[h]
 \centering \includegraphics[width=0.99\columnwidth]{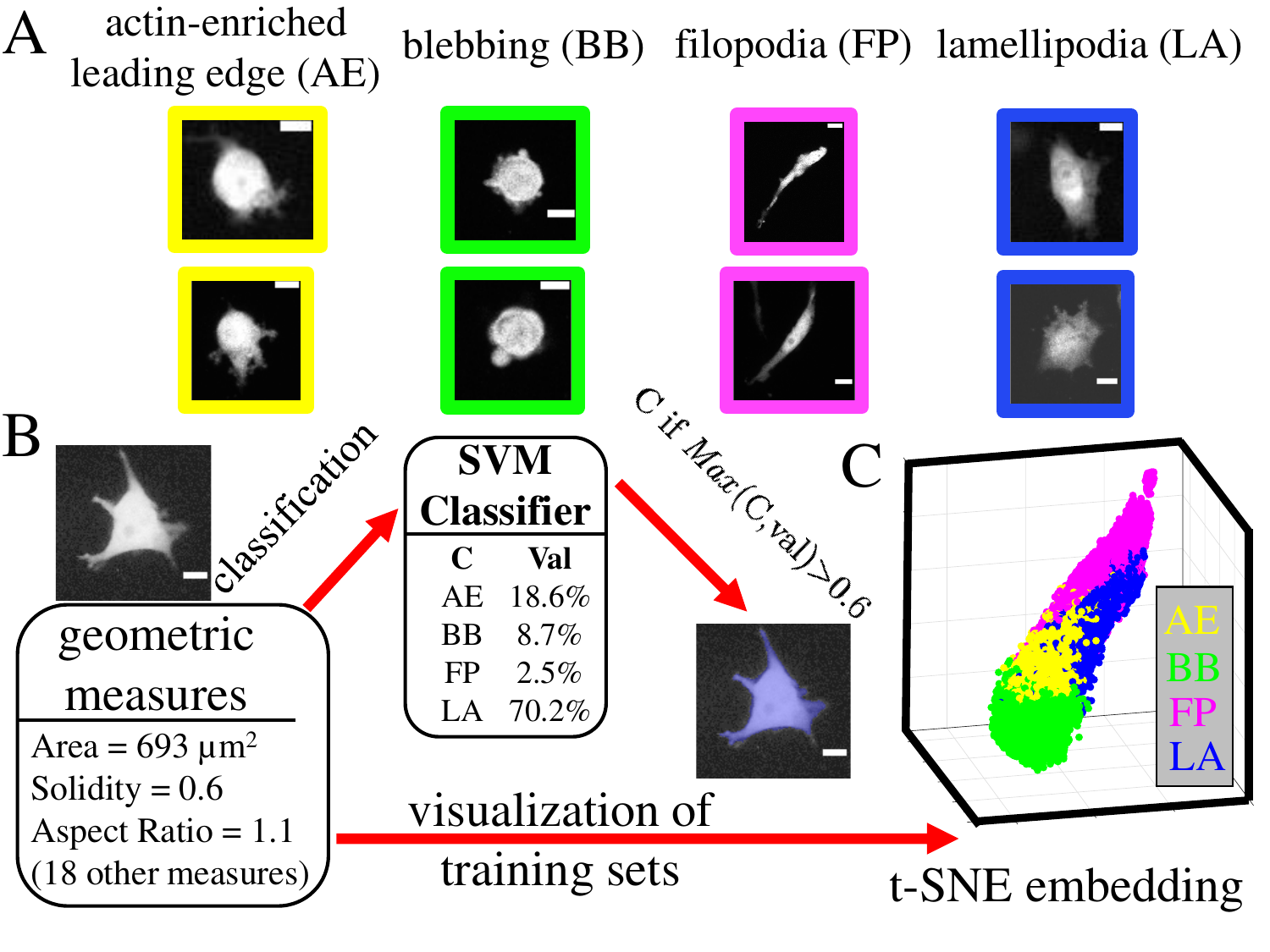}
  \caption{Development of a supervised machine learning model to classify cells into morphological phenotypes corresponding to different migration modes. (A) MDA-MB-231 cells in 3D collagen matrices exhibit multiple morphological phenotypes that are characteristic of four distinct migration modes: actin-enriched leading edge (AE), small blebbing (BB), filopodial (FP), and lamellipodial (LA). Scale bars are 20 $\mu$m. (B) The cell images are quantified using a total of 21 geometric measures such as area, solidity, and aspect ratio. With 3800 manually labeled single cell images we have trained a supported vector machine (SVM) to calculate probability scores (Val) for a cell to belong to each morphological phenotypes (classes). We assign a cell to the class $C$ with the maximum score ($Max(C, val)$), if this maximum score is greater than a threshold of 60\% ($Max(C, val)>$0.6). We consider a cell to be at an intermediate state if none of the four classes have a score higher than 0.6. In (B) a sample cell image is classified as a lamellipodial cell (LA), because LA class has a score of greater than 0.6. (C) To better visualize the high dimensional geometric measures, we apply t-SNE method to generate a 3D projection of the geometric cell shape space. 15,000 unseen data set is presented here. Different morphological phenotypes are well separated. AE (yellow): actin-enriched leading edge. BB (green): small blebbing. FP (magenta): filopodial. LA (blue): lamellipodial. This figure is prepared with Matlab R2020a (www.mathworks.com) and ImageJ (https://imagej.net).}
\label{fig2}
\end{figure}

After quantitatively demonstrating the shape fluctuations of migrating cancer cells, we next investigate cell morphodynamics at a mesoscale that allows us to gain insights on cell migration modes. This is possible because different migration modes are associated with distinct characteristic cell morphologies (Fig. \ref{fig2}A) \cite{Yamada2012commentary,Konstantopoulos2017}. Using 3800 manually labeled single cell images, we have trained machine classifiers that classify cell morphology into four morphological phenotypes based on and named after their corresponding migration modes. 

We consider four morphological phenotypes including two amoeboidal ones: actin-enriched leading edge (or AE in short) and small blebbing (BB); as well as two mesenchymal ones: filopodial (FP) and lamellipodial (LA). These morphological phenotypes are associated with known migration modes exhibiting characteristic molecular fingerprints. For instance AE and BB cells do not rely on cell-ECM adhesions during migration. For BB cells, their cortical stress continuously drives the formation of rounded blebs at the cell membrane \cite{BBref_charras2008blebs,BBref_lorentzen2011ezrin,BBref_yamazaki2005regulation}. AE cells, on the other hand, demonstrate elevated actin polymerization that drive sharp protrusions \cite{AEref_petrie2012leading,AEref_wyckoff2006rock}. Neither AE nor BB cells show clear polarization of cytoskeleton and cytoskeleton-associated proteins. In contrast, FP and LA cells exhibit strong cell-ECM adhesions. The filopodial cells consist of distinguishable F-actin bundles extending across the polarized cell body \cite{FPref_nalbant2004activation,trinkaus1973surface}, while the lamellipodial cells feature fan-shaped leading edges of migration \cite{LPref_abercrombie1970locomotion,LPref_petrie2012nonpolarized}. See also Fig. S19 for characteristic F-actin subcellular structures. Of note, another migration mode, namely lobopodial or nuclear piston mode, has not been observed in our experiments which is consistent with previous reports \cite{Yamada2017restore}. 

Once the classifier is trained, morphological phenotypes are determined automatically from a cell image if a particular phenotype receives more than 60\% probability score (Fig. \ref{fig2}B). For a small fraction of cells ($\approx$ 10\%), none of the four phenotypes receive more than 60\% probability score, we consider these cells to be in an intermediate state.

We have trained two classifiers (see \textit{SI Appendix} S4). The first one is based on support vector machines (SVM \cite{Cortes1995,Ben-Hur2001}) involving 21 geometric measures of binarized cell images. The second one is based on Random-Forest model using the same geometric properties \cite{Breiman2001}. The two classifiers agree with each other well on test data sets (90\% overlapping). The SVM classifier particularly has a higher success rate of classifying unseen data (88\%). In the following we mainly report the results from SVM algorithm. We also employ the t-SNE algorithm to reduce the dimension of the (21-dimensional) geometric shape space to facilitate visualization of cellular morphodynamics. As shown in Fig. \ref{fig2}C, unseen data (15,000 data points) belonging to different morphological phenotypes form separable clusters in the embedding space.

\begin{figure}[h]
 \centering \includegraphics[width=0.99\columnwidth]{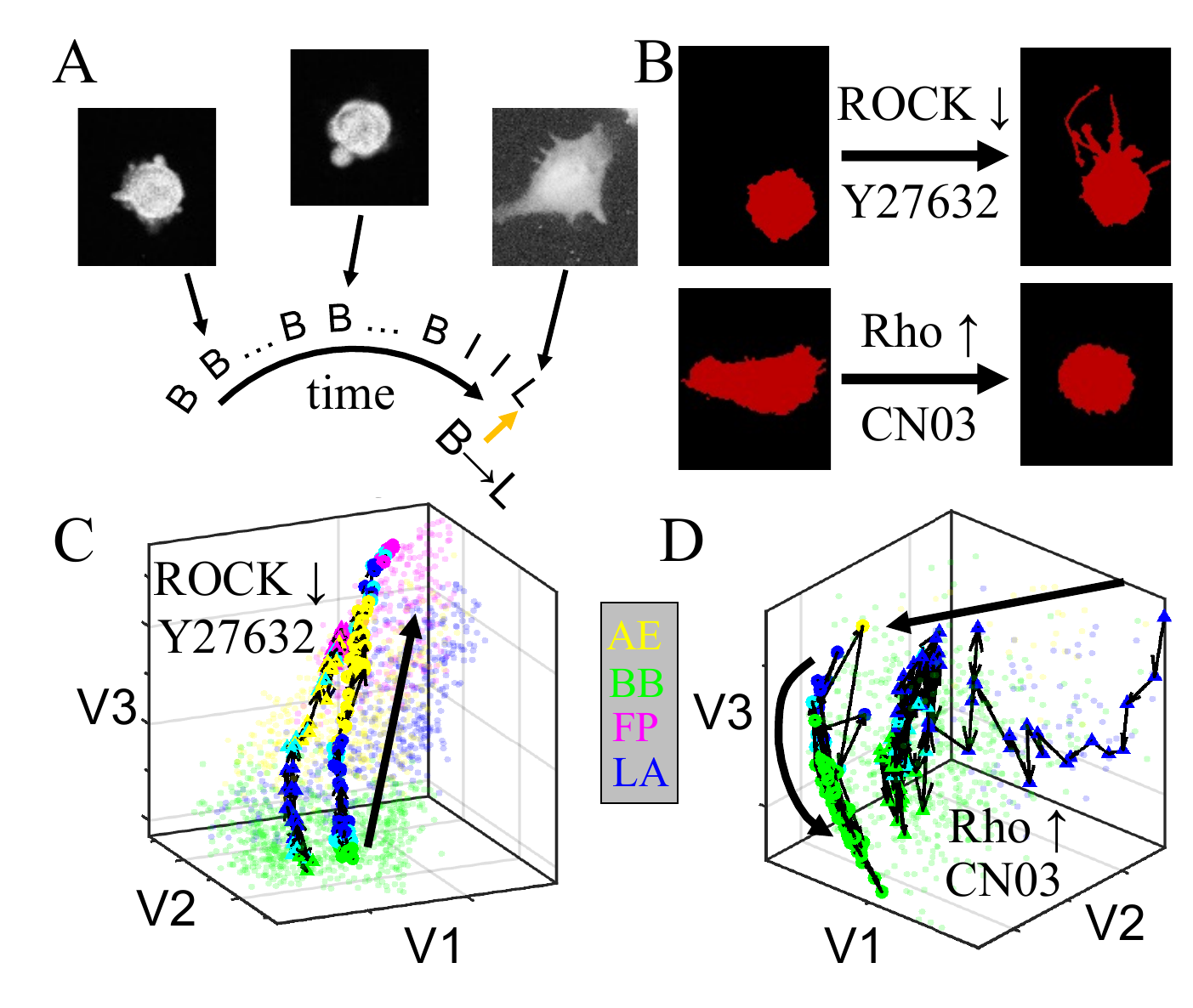}
\caption{Rho/ROCK-signaling internally controls mesoscale morphodynamics of 3D cultured MDA-MB-231 cells. (A) A sample time series of morphological phenotype. Insets: three snapshots showing the GFP-labeled cell morphology. Abbreviations: B -- blebbing, I -- intermediate state, L -- lamellipodial. (B) Representative morphological changes under treatment of Y27632 and CN03, which downregulates and upregulates Rho/ROCK signaling respectively. (C-D) Characteristic morphodynamic trajectories of cells in the t-SNE embedded shape space. The trajectories start immediately after introducing Y27632 or CN03, and ends after 12 hours of incubating with the drugs. The forward time directions are shown as thick curves with arrows as guide to the eyes. Two representative trajectories (one with circular symbols, and another one with triangular symbols) per each treatment are shown as colored symbols connected by black lines, where color represents the instantaneous phenotype. Unconnected light-colored dots show training sets which are the same as in Fig. \ref{fig2}C. Note that to better visualize the 3D trajectories, coordinates have been rotated with respect to  Fig. \ref{fig2}C. This figure is prepared with Matlab R2020a (www.mathworks.com) and ImageJ (https://imagej.net).}
\label{fig3}
\end{figure}

By applying the SVM classifier to time lapse recordings of 3D migrating MDA-MB-231 cells we find cells spontaneously make transitions among different morphological phenotypes. Fig. \ref{fig3}A shows snapshots of a typical cell. The cell switches directly from blebbing (B) mode to lamellipodial (L) mode via intermediate state (I). Therefore using machine learning technique we map cell morphodynamics into transitions between morphological phenotypes, or their associated migration modes.

In order to understand the mechanisms underlying cell morphological phenotype transitions, we examine the effects of manipulating Rho/ROCK-signaling, which is a master regulator that determines the mechanical state of a cell. Rho/ROCK-signaling controls key aspects of cell morphogenesis and migration, such as actomyosin contractility, actin polymerization, cell-cell and cell-ECM adhesion \cite{Schwartz2004_Rhoglacnce}. By regulating the cell mechanotransduction pathways, Rho/ROCK-signaling has also been shown to control 3D cell migration phenotype plasticity \cite{Marshall2005rho}.

We apply Y27632 \cite{Bissell2016_Y27632}, a potent Rho kinase (ROCK) inhibitor, and CN03 \cite{Isaacs2017_CN03}, a Rho-activator to MDA-MB-231 cells cultured in collagen ECM (see \textit{SI appendix} S5). Y27632 reduces actomyosin contractility, promoting transitions from blebbing to mesenchymal phenotypes \cite{Janmey2005}. On the other hand, CN03 elevates myosin II activity, leading to retraction of filopodia to rounded cell shapes (Fig. \ref{fig3}B). These results are consistent with previous reports on the molecular control of cell migration modes by Rho/ROCK-signaling \cite{Yamada2012commentary}.

While previous studies focus on the end points of manipulating Rho/ROCK-signaling, morphodynamic analysis offers insights to the transition paths between migration modes. In particular, we take advantage of a modified t-SNE algorithm \cite{Maaten2009}, which projects a cell image in the embedding space defined by the training sets (Fig. \ref{fig2}C, \textit{SI appendix} S4). This approach allows us to map the continuous shape change of a cell as a trajectory in the embedded shape space. Similar approaches have also been employed previously in studying complex body movements of other organisms such as fruit flies, where transition paths between different fly behaviors can be visualized \cite{Berman2014}.

Tracking the mesoscale morphodynamics of MDA-MB-231 cells under pharmacological perturbations, we find up and down regulation of Rho/ROCK-signaling do not lead to a reversal of morphodynamic trajectories. In particular, when treated with Y27632 blebbing cells turn to filopodial or lamellipodial via strongly converging trajectories most of which first visiting AE states (see also \textit{SI appendix} S5). Fig. \ref{fig3}C shows two representative trajectories. AE state exhibits weak cell-ECM adhesions and F-actin rich protrusions \cite{Soldatl2006,Sahai2006pseudopodia}. Our results suggest AE states mediate Rho/ROCK-signaling controlled transition from amoeboidal to mesenchymal motility. On the other hand, CN03 treatment causes the majority of mesenchymal cells to switch to blebbing modes. However, without going through AE states, CN03 leads to strongly fluctuating and diverging trajectories observed from multiple cells (Fig. \ref{fig3}D shows two representative trajectories, see also \textit{SI appendix} S5).

\begin{figure*}[t]
 \centering \includegraphics[width=1.8\columnwidth]{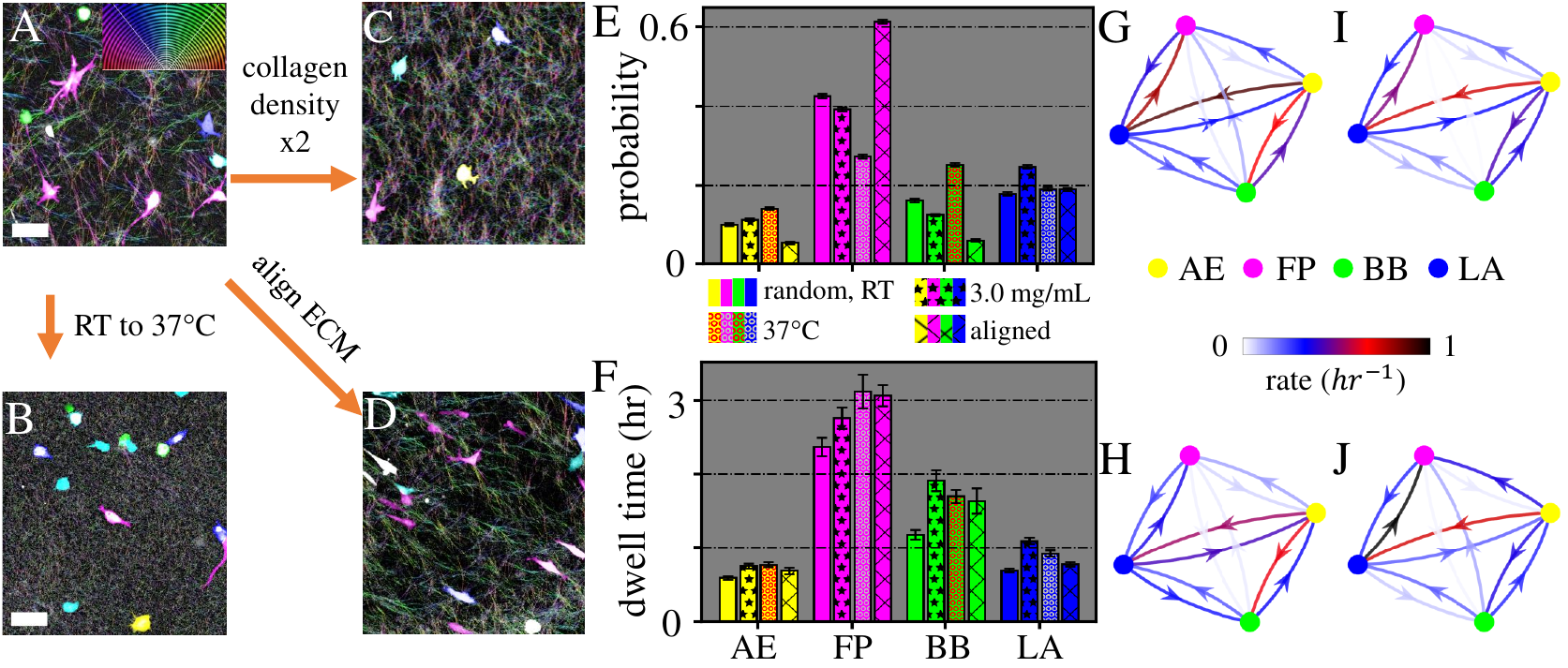}
  \caption{Physical properties of collagen ECM regulate the morphological phenotype homeostasis of 3D migrating MDA-MB-231 cells. (A-D) Confocal reflection images and pseudo colored MDA-MB-231 cells for collagen matrices prepared at varying conditions. Scale bars: 20 $\mu$m. A: collagen ECM prepared at room temperature (RT, or 25 $^\circ$C) and collagen concentration of $[col] = $ 1.5 mg/mL. B: collagen ECM prepared at 37 $^\circ$C and [col] = 1.5 mg/mL. C: collagen ECM prepared at RT and $[col] =$ 3.0 mg/mL. D: collagen ECM prepared with flow-aligned collagen fibers. (E) Fraction of cells in each morphological phenotype. 8,000 single cell images are analyzed under each ECM condition. (F) Dwell time of cells in each morphological phenotype. Errorbars in (E-F) represent 95\% confidence intervals calculated from 1000 bootstrap iterations. (G-J): The transition matrix -- morphological phenotype transition rates under varying ECM conditions. G: collagen ECM prepared at room temperature and $[col] = $ 1.5 mg/mL. H: collagen ECM prepared at 37 $^\circ$C and $[col] =$ 1.5 mg/mL. I: collagen ECM prepared at RT and [col] = 3.0 mg/mL. J: collagen ECM prepared with flow-aligned collagen fibers. Under each ECM condition a total of more than 2,000 hours of single cell trajectories are analyzed. This figure is prepared with Matlab R2020a (www.mathworks.com) and ImageJ (https://imagej.net).}
\label{fig4}
\end{figure*}

Since Rho/ROCK-signaling is employed by cells to sense ECM physical properties, we next investigate the external control of mesoscale cell morphodynamics. In particular, we focus on the role of ECM physical properties in regulating cell morphological phenotype transitions. In order to control the microstructure of collagen matrices we apply three methods as shown in Fig. \ref{fig4}(A-D) (see also \textit{SI appendix} S6). First, increasing gelation temperature from room temperature (RT) to 37 $^\circ$C, while keeping collagen density at [col]= 1.5 mg/mL significantly reduces fiber length and pore size (Fig. \ref{fig4}B). Second, increasing collagen density to [col]= 3.0 mg/mL while keeping gelation temperature at room temperature moderately reduces pore size, preserves a clear fibrous structure, and increases stiffness (Fig. \ref{fig4}C). Finally, keeping gelation at room temperature and [col]=1.5 mg/mL, while generating an unidirectional flow field during gelation leads to aligned collagen fibers in the ECM. This method creates strong anisotropy in the ECM. 

We find the occurrence probability (population fraction) of different morphological phenotypes are remarkably different at different ECM conditions. As shown in Fig. \ref{fig4}E, increasing gelation temperature does not affect the probability of AE and LA cells. However, the homogeneous matrix microstructure at 37 $^\circ$C significantly reduces the fraction of FP cells from 43\% to 25\%, while increases fraction of BB cells from 15\% to 25\%. Compared with increasing gelation temperature, doubling collagen concentration leads to less dramatic changes of the ECM microstructure. Correspondingly, only moderate changes of phenotype probabilities are observed. On the other hand, when matrix anisotropy is increased by aligning ECM fibers, we find significant shift of cells from amoeboid phenotypes to filopodial mode. Taken together, these results show that ECM heterogeneity and anisotropy determine the probability of different morphological phenotypes.

We have also examined the stability of each morphological phenotype by measuring the average dwell time -- duration of a cell to stay continuously in a morphological phenotype before transition to another (\textit{SI appendix} S7). As shown in Fig. \ref{fig4}F, in all three cases manipulating ECM physical properties moderately increase the dwell time of all four morphological phenotypes. Therefore the changes in the phenotype probability can not be explained by the phenotype stability alone, and in some cases move in opposite trend from the dwell times observed. For instance, occurrence probability of filopodial cells is lowest at 37 $^\circ$C, which happens to be the condition where filopodial cells show longest dwell time. As we will show later, such discrepancies can be attributed to the detailed transition paths between the phenotypes.

To reveal further details of morphological phenotype dynamics, we have computed the phenotype transition matrix: rates $r$ that characterize the probability of transitions per hour between any two phenotypes (Fig. \ref{fig4}G-J (see also \textit{SI appendix} S7). While the rates vary dramatically for different ECM conditions (arrows in Fig. \ref{fig4}G-J), we notice several remarkable common features. First, direct transitions along FP - BB path rarely happen ($r<$0.03 hr$^{-1}$). Instead, amoeboidal - mesenchymal transitions are primarily mediated by LA and AE states, presumably by turning cell-ECM adhesion on and off. On the other hand, transitions within the amoeboidal (AE - BB) and mesenchymal (FP - LA) modes are frequent, and the rates can go up to 1 per hour. Finally, while the morphological phenotype transitions are intrinsically non-equilibrium processes, probability fluxes between states are generally very small (\textit{SI appendix} S7). This means that an approximate detailed balance exists among morphological phenotypes. In comparison with other nonequilibium stationary processes at mesoscale \cite{MacKintosh2016_flux}, we speculate morphological phenotype transitions are not gated by active processes such as ATP consumption.

The transition rates also offer insights to understand the ECM-dependence of the fraction of cells in each morphological phenotype (Fig. \ref{fig4}E). For instance, as gelation temperature increases from RT to 37 $^\circ$C, rates from AE to FP decreases by 52 percent, and rates from BB to AE decreases by 22 percent (Fig. \ref{fig4}G and Fig. \ref{fig4}H). As a result, we observe more blebbing cells and less filopodial cells in collagen matrices prepared at 37 $^\circ$C. This is consistent with the mechanical mechanism of blebbing formation \cite{Tinevez2009,Yamada2012nonpoloarized}. Blebs form when actomyosin contractility exceeds the binding between cortical actin and cell membrane. A blebbing cell turns to AE when actin polymerization causes sharp protrusion on the membrane. Our results suggest that homogeneous collagen ECM favors BB phenotype to AE, likely due to the reduced protrusive force associated with actin polymerization.
 
Conversely, as ECM becomes more anisotropic (Fig. \ref{fig4}G and Fig. \ref{fig4}J), the transition rate from LA to FP increases as much as 27 percent, while rate from AE to BB decrease by 44 percent. Together, these altered rates lead to a significant fraction of blebbing cells turning to filopodial as shown in Fig. \ref{fig4}{E}. Filopodial protrusions consist of elongated F-actin bundles supported by elevated actin polymerization and cross-linking by Ena/VASP proteins \cite{Borisy2004}. Our results suggest that the mechanical barrier separating filopodia and blebbing protrusions is too high for actomyosin contractility to overcome directly. Instead, a blebbing cell turning into a filopodial one has to first transform into AE or LA states.

\begin{figure}[h]
 \centering \includegraphics[width=0.99\columnwidth]{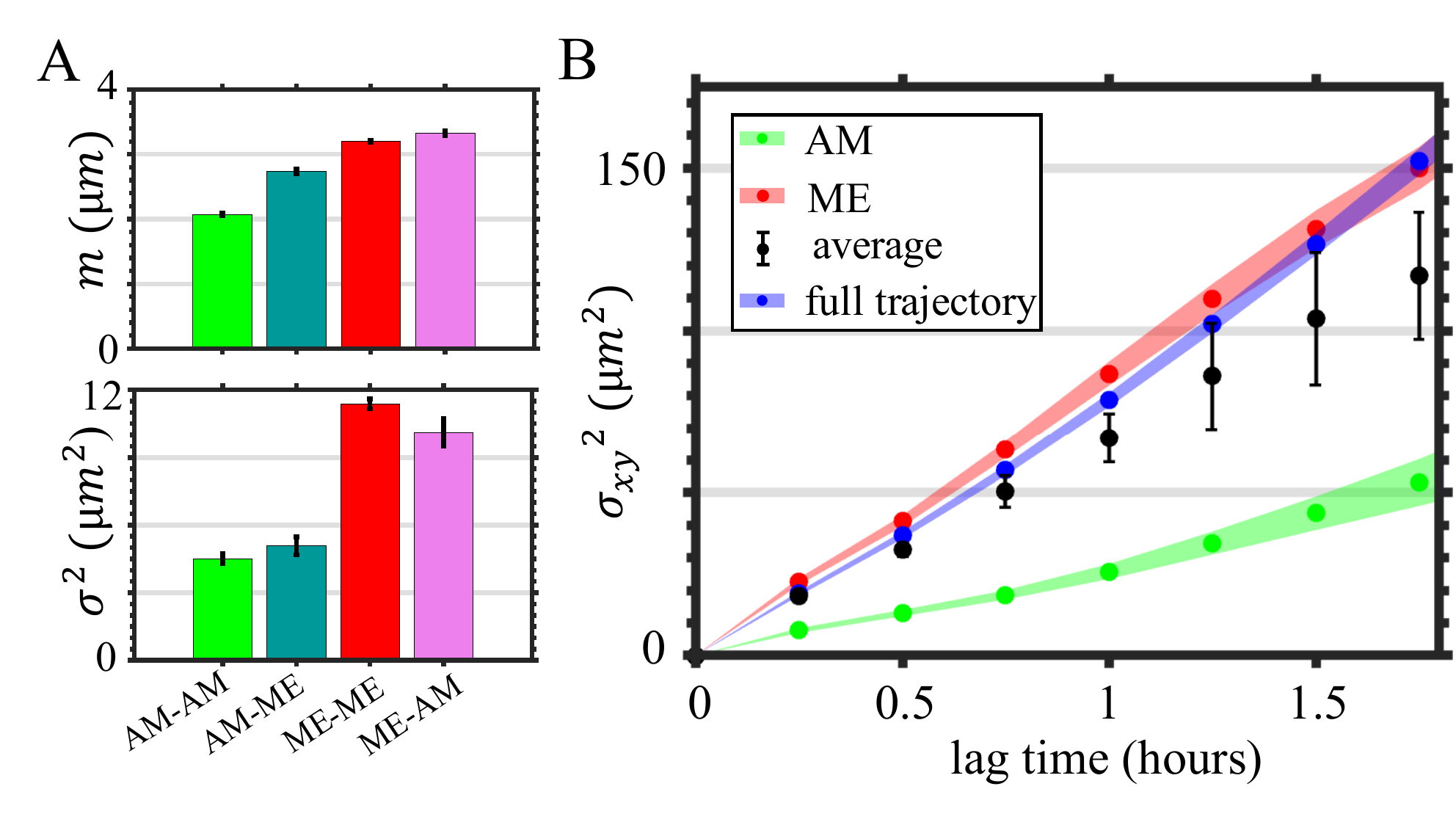}
 \caption{Cancer cell migration in 3D ECM is coupled with morphological phenotype and phenotype transitions. (A) Means ($m$, upper panel) and variances ($\sigma^2$, lower panel) of the step size distributions by fitting the experimental measurements with log-normal distribution. The steps are categorized based on morphological phenotype (coarse-grained as mesenchymal, amoeboidal and intermediate state) dynamics. ME: abbreviation for mesenchymal, which includes FP and LA states. AM: abbreviation for amoeboidal, which includes BB and AE states. If a cell makes a step by starting from AM state to ME state, the step is categorized as an AM-ME step. The starting and ending frames of steps are separated by 15 minutes. In (A)  errorbars show the 95\% confidence interval of fitted parameters. (B) Real-space (2D projection) mean square displacements (MSD) of cells. AM: the MSD of cells dwelling in the AM state. ME: the MSD of cells dwelling in the ME state. average: the weighted average of mean square displacements of AM, ME, and intermediate dwelling cells. The average is based on the occurrence fraction of AM, ME and intermediate state of cells. Full trajectory: the MSD obtained from entire cell trajectories regardless of the morphological phenotypes. Shaded areas in C show SEM. In (A-B) a total of 1974 hours of single cell trajectories are analyzed. The ECM are prepared at room-temperature with a concentration [col] = 1.5 mg/mL. This figure is prepared with Mathlab R2020a (www.mathworks.com) and ImageJ (https://imagej.net).}
\label{fig5}
\end{figure}

Because the morphological phenotype of a cell is linked to its 3D migration mode, we next investigate if the invasion potential of MDA-MB-231 cells depends on the mesoscale morphodynamics. Due to the short dwell times for each morphological phenotype, we only consider two coarse-grained classes of morphologies: mesenchymal (ME), which consists of FP and LA states; and amoeboidal (AM), which consists of AE and BB states. In particular, we measure for short time scales the step size distributions and for longer time scales the mean square displacement of the cells in randomly aligned collagen matrices gelled at room temperature (Fig. \ref{fig5}).

Interestingly we find the steps are better described by a log-normal, rather than Gaussian distribution (\textit{SI appendix} S8) due to frequent large steps. Fig. \ref{fig5}A shows the mean and variance of the fitting parameters. It is clear that the steps in physical space are coupled with the corresponding mesoscale dynamics. For cells that dwell in the amoeboidal class, both mean and variance of the steps are the smallest. Correspondingly, the mean square displacement of amoeboid cells have a small slope, corresponding to an effective diffusivity of 6 $\mu$m$^2$/hour (for each spatial dimension, Fig. \ref{fig5}B). On the other hand, cells make larger steps when dwelling in the mesenchymal class, and the effective diffusivity increases by three-fold to 19 $\mu$m$^2$/hour. The observed higher motility for cells in the ME state than in AE state is consistent with both migratory measurements \textit{in vitro} \cite{Shamik2017} and metastasis measurements \textit{in vivo} \cite{Pugacheva2017}. For instance, using a mouse breast cancer model, it is shown that inhibition of mesenchymal phenotype by NEDD9-depletion significantly (by $\approx$ 50\%) reduces the number of circulating tumor cells.  

Our analysis shows that not only it is important to distinguish different morphological phenotypes in studying the motility of cancer cells, but also one may need to take into account of phenotype transitions. We find cell migration steps associated with different class-switching events have distinct statistical distributions (Fig. \ref{fig5}A, also see SI appendix S8 for the motility characteristics of cells involving intermediate states). Without accounting for the class-switching events, the weighted average of mean square displacements from mesenchymal, amoeboidal, and intermediate state cells underestimates the observed cell motility by 20\% (Fig. \ref{fig5}B, the weighted average MSD curve corresponds to a diffusivity of approximately 15 $\mu$m$^2$/hour, as compared with 19 $\mu$m$^2$/hour for full cell trajectories). We also find cell motility depend on the direction of state transitions. For instance, the variance of steps coupled to ME-to-AM transition is almost twice the value for AM-to-ME transition (Fig. \ref{fig5}A, lower panel). Since morphological phenotype transitions occur spontaneously at single cell level, our results show that morphodynamics and cell motility are closely coupled to determine the invasive potential of cancer cells. 

\begin{figure*}[t]
 \centering \includegraphics[width=1.8\columnwidth]{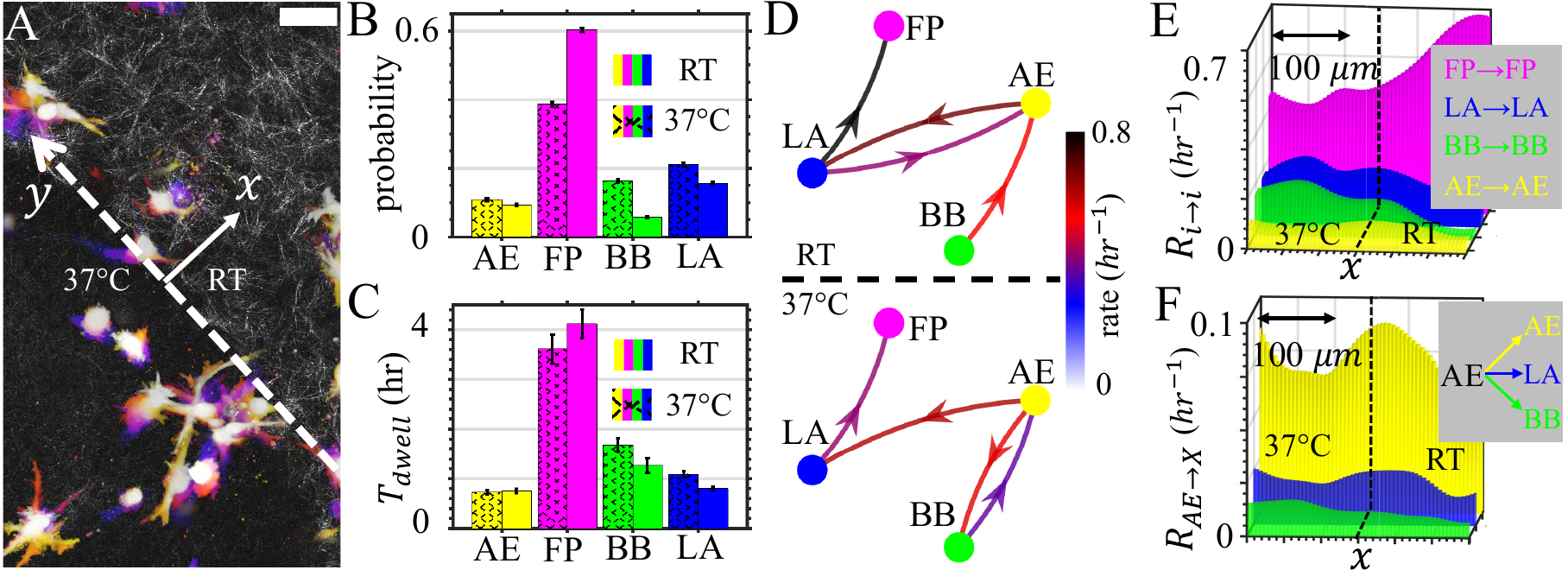}
 \caption{Morphological phenotype transition facilitates cell migration in heterogeneous ECM. (A) Time-lapse projection of 3D migrating MDA-MB-231 cells navigating engineered heterogeneous ECM. The ECM contains two adjacent layers that are prepared at room temperature (RT) and 37 $^\circ$C respectively. A confocal reflection image shows the ECM structure next to the interface (dashed line, $y$-axis). Scale bar: 50 $\mu$m. (B) Fraction of cells of each morphological phenotype in both sides of the interface. (C) Dwell time of cells of each morphological phenotype in both sides of the interface. (D) Phenotype transition rates in both sides of the interface. (E) Spatial frequency of of dwell events. (F) Spatial frequency of AE to AE, AE to LA and AE to BB events. See main text for the definition of spatial frequency $R(i \rightarrow j,x)$. In (E-F) dashed lines indicate the interface ($x$=0) separating the 37 $^\circ$C gel (left) and RT gel (right). A total of 3,800 hours of single cell recordings from three independent experiments have been used to calculate the results in (B-F). This figure is prepared with Mathlab R2020a (www.mathworks.com) and ImageJ (https://imagej.net).}
\label{fig6}
\end{figure*}

During metastasis a migrating cancer cell must navigate ECM of distinct mechanical properties. Therefore, we next investigate how cell morphodynamics facilitate cell traverse interfaces and adapt to ECM of distinct mechanics. To this end, we create collagen matrices consist of two integrated layers (\textit{SI appendix} S9). The RT layer is prepared at room temperature that shows a porous fibrous network, and the 37 $^\circ$C layer is prepared at 37 $^\circ$C showing a much more homogeneous structure (Fig. \ref{fig6}A). Without additional cues MDA-MB-231 cells randomly navigate the ECM, occasionally traverse the interface to experience a sudden change of ECM physical properties. Over the course of 24 hours we do not observe durotaxis. 

Consistent with the corresponding results in uniform ECM, we find the likelihood of observing a filopodial cell is significantly higher in the ECM layer prepared at room temperature, while for blebbing cells the probability is higher in the gel layer prepared at 37 $^\circ$C (Fig. \ref{fig6}B). The dwell times of phenotypes, on the other hand, follow the same trend of occurrence probabilities but change rather moderately (Fig. \ref{fig6}C). 

The shift of phenotype homeostasis once again can be understood from the phenotype transition matrices. To simplify the discussion in Fig. \ref{fig6}D we plot the top four matrix elements of the transition matrices calculated from cells in each of the two layers. In the RT layer, filopodial cell population is enriched by frequent LA-AE exchange and LA to FP transition. As cells cross the interface, LA to FP becomes less likely, and the AE-BB pathway is steered to favor blebbing states. 

To further quantify the effect of the interface in modulating cell morphodynamics, we calculate spatial frequencies of dwell events (Fig. \ref{fig6}E) and AE-originating transition events (Fig. \ref{fig6}F). We define a coordinate system where the y-axis is along the interface passing $x$=0 (Fig. \ref{fig6}A). This allows us to combine data from multiple repeating experiments where cell locations are seeded randomly. After aligning the coordinates, we define the spatial frequency of transition (or dwell) events from state $i$ to state $j$ as $R(i \rightarrow j,x)$, which satisfies
\begin{eqnarray}
P(i \rightarrow j,x) = R(i \rightarrow j,x)M(x)
\end{eqnarray}
Here $P(i \rightarrow j,x)$ is the probability density of observing event $i \rightarrow j$ per unit time (1 hour), and $M(x)$ is the cell density (along $x$-axis). We use a 1-D Gaussian kernel to estimate $P(i \rightarrow j,x)$ and $M(x)$ (\textit{SI appendix} S9). As a result, the spatial frequency $R(i \rightarrow j,x)$ represents the likelihood of a cell to undergo a specific type of transition over unit time (1 hour) as a function of distance to the interface.

The spatial frequency of dwell events clearly show that while FP state is increasingly stable into the RT layer, LA and BB states are more stable in the 37 $^\circ$C layer. AE state, on the other hand, is most stable at the interface (Fig. \ref{fig6}E). Therefore AE state plays a special role in mediating the cell adaptation across distinct ECM layers. Indeed, we find a gradual shift of favorable AE-originating transitions as distance to the interface varies. The frequency of AE to LA events, the main amoeboidal to mesenchymal path, peaks in the RT layer. AE to BB events, which is mainly responsible of enriching blebbing cells, has peak frequency in the 37 $^\circ$C layer. Taken together, we find morphological phenotype transitions and the associated migration mode switching are integral parts of cancer cell invasion and adaptation to complex ECM.

\section{Discussion}
\label{sec:discuss}
In this paper, we report the morphodynamics of MDA-MB-231 cells in type I collagen ECM as a model system of metastatic cancer cells migrating in 3D tissue. MDA-MB-231 cells rapidly change their geometry, exhibiting a subdiffusive random walk in the geometric shape space. This occurs simultaneously with their superdiffusive walks in the real space (Fig. \ref{fig1}). 

The biological significance of the morphodynamics is further demonstrated by classifying cell shapes into morphological phenotypes corresponding to different migration programs (Fig. \ref{fig2}). This allows us to study cell morphodynamics at the mesoscale, in terms of morphological phenotype transitions. Utilizing machine learning and visualization techniques, we show that cell morphodynamics is regulated by Rho/ROCK-signaling (Fig. \ref{fig3}), which is a molecular control hub of cell mechanosensing and force generation. It has been shown previously that Rho/Rac signaling regulates the shift between mesenchymal and amoeboidal motility \cite{Marshall2008rac,Marshall2010plasticity}. Our analysis further reveals that instead of favoring a particular mode of motility, perturbations of Rho/ROCK-signaling alter the migration mode transition rates. In particular, down regulating Rho leads to overall amoeboidal-to-mesenchymal transition that routes through AE and LA states. Activation of Rho, on the other hand, leads to strongly fluctuating morphodynamics that enriches blebbing cells. The irreversibility of up and down regulating Rho/ROCK-signaling results suggest a complex phenotype landscape that controls 3D cancer cell motility.

We study morphological phenotype transitions in ECM of distinct physical properties and find ECM microstructure modulates the probabilities, dwell times, and transition rates of morphological phenotypes. Collagen matrices with homogeneous structure, as those prepared at higher temperature, enrich the population of blebbing cells. By comparing the transition matrices, we find the enrichment of blebbing cells is directly related with the reduced transition rate from BB to AE state, and also indirectly contributed by the mesenchymal-to-amoeboidal transition through LA and AE states. Similarly, collagen matrices with structural anisotropy enrich the population of filopodial cells. The enrichment is directly attributed to an increased LA to FP rate, and indirectly contributed by the amoeboidal-to-mesenchymal transition mediated by LA and AE states. These results show that it is possible to execute external control of cell morphodynamics (and the corresponding 3D migration modes) through ECM mechanics. Importantly, taking into account of the phenotype transitions allows us to better predict the outcome of manipulating cell migration mode through ECM physical properties \cite{Yamada2016adhesion,Konstantopoulos2017}.

In light of the rapid phenotype transitions exhibited by individual cells, 3D cancer cell motility may be considered as a hidden Markov process where each phenotype is associated with characteristic step size distributions (Fig. \ref{fig5}). Specifically, we find steps that occur simultaneously with a phenotype transition have distinct sizes compared with steps that occur while cells dwell in a particular morphological phenotype. This makes morphodynamics a crucial factor in determining the invasive potential of cancer cells. To our knowledge, this aspect has been so far largely overlooked in the literature.

In the lens of a hidden Markov process, morphodynamics may facilitate cancer invasion because phenotype transitions allow cancer cells to search for and commit to a more effective migration program \cite{Lander2011}. Using a ECM model consisting of two mechanically distinct layers, we show the cells gradually adjust their morphodynamics as they approach and cross the layer interface (Fig. \ref{fig6}). Therefore morphological phenotype transitions may be essential in cancer cell metastasis by enabling the cells to navigate non-uniform ECM.

The connection between morphological phenotypes and cell migration modes may be further strengthened by incorporating key molecular fingerprints. Using a small data set, we find the machine-classified morphological phenotypes have F-actin structures that are expected from the corresponding migration modes (Fig. S19, \cite{trinkaus1973surface}). We anticipate that training with a multichannel data set which includes not only cell shape but also molecular profiles such as actin, myosin, and integrin, will improve accuracy in identifying cell migration modes. However, doing so could reduce the throughput and limit the applicability in general imaging settings.

In summary, we demonstrate the morphodynamics of 3D migrating cancer cells as a powerful tool to inspect the internal state and microenvironment of the cells. Investigated at mesoscale, the morphodynamics imply that 3D cancer cell migration is inherently plastic \cite{Yamada2012commentary}. The plasticity is controlled by the mode transition matrices, rather than a deterministic decision tree \cite{Konstantopoulos2017}. In order to further exploit the information provided by the cell shape fluctuations, future research is needed to decode morphodynamics as a rich body language of cells, and to control morphodynamics as a route of mechanical programming of cell phenotype.

\section{Materials and methods}
\label{sec:methods}
GFP-labeled MDA-MB-231 cells are purchased from GenTarget Inc and are handled following the vendor's recommendations. High concentration rat tail Type I collagen solutions are purchased from Corning Inc, and the collagen gels are prepared following standard protocol. Cells and collagen fibers are imaged using Leica SPE confocal microscope with fluorescent and reflection channels simultaneously. A streamline of utilizing ImageJ , Matlab, and Python scripts are developed to analyze raw images. See \textit{SI Appendix} S1-S10 for details of 3D cell culture, microscopy, pharmacological treatments, and data analysis.

%\section{acknowledgments}
\begin{acknowledgments}
This is a preprint of an article published in Scientific Reports. The final authenticated version is available online at: https://doi.org/10.1038/s41598-021-99902-9. We thank Prof. Michelle Digman and Prof. Steve Press{\'e} for helpful discussions. We also thank Prof. Willie Rochefort and Conor Harris for help with rheological measurements. The funding for this research results from a Scialog Program sponsored jointly by Research Corporation for Science Advancement and the Gordon and Betty Moore Foundation through a grant to Oregon State University by the Gordon and Betty Moore Foundation (award 6790.11). Part of this research was conducted at the Northwest Nanotechnology Infrastructure, a National Nanotechnology Coordinated Infrastructure site at Oregon State University which is supported in part by the National Science Foundation (grant NNCI-1542101) and Oregon State University. C. Eddy and B. Sun are supported by DOD award W81XWH-20-1-0444 (BC190068). B. Sun is also supported by the National Institute of General Medical Sciences award 1R35GM138179 and National Science Foundation award PHY-1844627.

\end{acknowledgments}

%% Enter the largest bibliography number in the facing curly brackets
%% following \begin{thebibliography}

\bibliography{morphodynamics_arxiv}

%%%%%%%%%%%%%%%%%%%%%%%%%%%%%%%%%%%%%%%%%%%%%%%%%%%%%%%%%%%%%%%%

%% Adding Figure and Table References
%% Be sure to add figures and tables after \end{article}
%% and before \end{document}

%% For figures, put the caption below the illustration.
%%
%% \begin{figure}
%% \caption{Almost Sharp Front}\label{afoto}
%% \end{figure}

%% For Tables, put caption above table
%%
%% Table caption should start with a capital letter, continue with lower case
%% and not have a period at the end
%% Using @{\vrule height ?? depth ?? width0pt} in the tabular preamble will
%% keep that much space between every line in the table.

%% \begin{table}
%% \caption{Repeat length of longer allele by age of onset class}
%% \begin{tabular}{@{\vrule height 10.5pt depth4pt  width0pt}lrcccc}
%% table text
%% \end{tabular}
%% \end{table}

%% For two column figures and tables, use the following:

%% \begin{figure*}
%% \caption{Almost Sharp Front}\label{afoto}
%% \end{figure*}

%% \begin{table*}
%% \caption{Repeat length of longer allele by age of onset class}
%% \begin{tabular}{ccc}
%% table text
%% \end{tabular}
%% \end{table*}

\end{document}